%% file: main.tex
\documentclass[letterpaper,twocolumn,10pt]{article}
\usepackage{usenix,epsfig,endnotes}
\usepackage{hyperref}

\usepackage[table,xcdraw]{xcolor}
\usepackage{amsmath,amsfonts}
\usepackage{graphicx,color} %
\usepackage{textcomp}
\usepackage{soul} %
\usepackage{booktabs} %
\usepackage{array}
\usepackage{multirow}
\usepackage{subfig}
\usepackage[inline]{enumitem} %
\PassOptionsToPackage{hyphens}{url} %
\usepackage{url}
\usepackage{xcolor}
\usepackage{color,colortbl}
\usepackage{tcolorbox}
\usepackage{makecell}
\usepackage[font=small,skip=0pt]{caption}

\newcommand{\secref}[1]{\S \ref{#1}}

\newcommand{\leo}[1]{election official{#1}}
\newcommand{\Leo}[1]{Election official{#1}}

\begin{document}

\date{}

\title{\Large \bf ``Bless his heart... he thought all we did was push a button'': \\ Understanding Worker Challenges with U.S. Election Technology}

\author{
{\rm Delaney Gomen}\\
Georgia Institute of Technology 
\and
{\rm Josiah Hester}\\
Georgia Institute of Technology\\
\and
{\rm Naveena Karusala}\\
Georgia Institute of Technology\\
\and
{\rm Michael Specter}\\
Georgia Institute of Technology\\
}

\maketitle

\thispagestyle{empty}

\subsection*{Abstract}
This paper explores how the technologies U.S. election officials depend on also challenge their work. Every election, misleading narratives stem from errors that occur while using election technology, threatening election official safety and eroding faith in elections. We seek to understand the impact of these challenges directly from worker perspectives. This paper presents a qualitative analysis using combined interview and survey data from a total of 50 election officials representing 20 states. Our findings present design as a substantial factor in mistakes when using election technology, but other obstacles, like lack of funding and technical support, also strain operations. We show how election technology design can lead to harms for election officials and connect those harms to their negative effects on democracy. Our work emphasizes the potential for researchers in human-centered computing to support U.S. election officials---and democracy---by working towards better usability across election technologies. 

\input{1introduction}
\input{2background}

\input{3relatedwork}
\input{4methods}

\input{5findings}
\input{6discussion}

\input{7conclusion}

\section*{Acknowledgments}
We would like to thank our participants who volunteered their time to this study. We would like to thank Matthew Bernhard, Veronica Rivera, and Joe Kirk for their thoughtful input on drafts of this work. 

{\footnotesize \bibliographystyle{acm}
\bibliography{references}}

\input{8appendix}

\end{document}

%% file: 1introduction.tex
\section{Introduction}

\begin{quote}
    ``I had one observer come in probably six or seven years ago. And bless his heart, he sat there election night and watched me upload everything and walked out and told me, he thought all we did was push a button... No matter how much I say to [constituents], you know, come and see what we're doing. Come election day, come during early voting, watch us set up the machines and program them. They don't consider it important.'' \textit{---P03}
\end{quote}

In the U.S., election officials\footnote{Workers employed by local governments in the U.S. to conduct elections. See \secref{background:electionwork} for definitions of work used in this paper.} manage an array of bespoke election technology. Voter registration databases are used daily to keep track of eligible voters. Electronic pollbooks, tabulators and scanners, and more are deployed across an entire jurisdiction for multiple election days every year. These operations happen with tight budgets and limited staff in an increasingly dangerous political environment---in the last several years, election officials and poll workers in the U.S. have been victims of physical and digital violence at an unprecedented rate  \cite{associatedpressFailedNewMexico2023, arnold_for_2023, vasilogambros_racist_2024, judiciary_protecting_testimony, edelmanadamMotherdaughterElectionWorkers2022}. However, little is known about \textit{how} election officials interact with election technology. Do these technologies meet the needs of election officials, and if not, what changes are needed to properly support election work? 

These questions are especially important to answer given the impact election technology errors have on democracy. Even minor incidents that are resolved can fuel election misinformation and disinformation. Starbird et al. described election issues as ``fodder for false rumors and misleading narratives'' \cite{WhatExpectWhen2024}. The Election Integrity Partnership found that 49\% of the misinformation reports it processed during the 2020 election involved an ``exaggerated issue''  \cite{observatoryLongFuseMisinformation}. Prior work that traces election misinformation to its origins show that election technology specifically is at the center of many false and exaggerated claims \cite{starbirdWhatGoingEvidenceframe2025, observatoryLongFuseMisinformation, prochaskaDeepStorytellingCollective2025}. For example, in Antrim County, Michigan, an erroneous vote count in 2020 became a central story of viral conspiracy theories that were absorbed into \#StopTheSteal, a movement that culminated into the January 6, 2021 insurrection \cite{MichigansAntrimCounty, feuerJan6Story2022}. A 2022 report of the incident conducted by an election security expert determined the cause was human errors compounded by ``insufficiently defensive software design'' \cite{haldermanAntrimCounty20202022}. This author also noted there has been ``little attention to usability problems confronting election workers'' and called on researchers to address this issue \cite{haldermanAntrimCounty20202022}. 

While Human-Computer Interaction (HCI) researchers have long studied the usability of election technology, this work largely focuses on \textit{voters}, not election officials or poll workers, and many focus on an isolated election technology \cite{olemboEVotingSystemUsability2013, acemyanAssessingUsabilityHart2017a, bernhardCanVotersDetect, bedersonElectronicVotingSystem2003, robertsonVotercenteredDesignVoter2005, selkerVotingUserExperience2003, markyImprovingUsabilityUX2020, hiltWhyJohnnyChecks2026}. Even though these past findings are relevant for future assessments of worker-facing technologies, there could be unique challenges that only manifest when multiple components interact in active elections. Confusing buttons that led to an error in Milwaukee would likely be captured in a focused usability study \cite{bayatpourCBS58Cameras}, but the design choices that led to election workers in Georgia accidentally double-scanning ballots and then not being able to determine which had been double-scanned are less obvious to identify a priori \cite{niesse_fulton_2024}. In Pennsylvania, clicking on the wrong file caused a printing issue that affected voting, an issue not detectable by analyzing any election technology individually \cite{HumanErrorChester2026}. In this high-stakes environment, capturing such edge-cases like \cite{staff_voting_2023, harrington_election_2022, parksFloridaGovernorSays2019, quinlanColoradoVotingSystem2024} is crucial. With no formal public review process to analyze these incidents holistically, there is a need for HCI researchers to study U.S. election technology in the context of election work.

In order to understand challenges election officials face when using election technology and address these problems, we analyze data from interviews with 14 election officials along with 45 survey responses from election officials representing 20 U.S. states. Our interview and survey questions were designed to answer our research questions: 

\begin{itemize}
    \item \textbf{RQ1:} What challenges do U.S. election officials face when leveraging technology in election work? How do these challenges impact workers and U.S. elections? 
    \item \textbf{RQ2:} How do U.S. election officials develop, share, and find solutions to overcoming these challenges?

\end{itemize}

In our findings, we find election technology design is the crux of many reported issues, but funding, accessibility, and reliability also impact technology use. We document how these issues turn into consequences for election workers and elections. For example, election officials anticipate confusion and public scrutiny when election technology fails. That scrutiny turns into safety risk for election workers via misinformation and threats from the public \cite{parksMostDetailedLook2024, parks15Local2022}, which then impacts how they approach resolving human error and security weaknesses in the future. These threats negatively affect election worker retention \cite{parks_voting_2025}, leading to shortages in staff. Our participants share how those shortages lead to more mistakes and, therefore, more potential for misinformation---a cycle of democratic harm.  

This paper makes three main contributions to the HCI community. First, we provide an exploratory analysis of how U.S. election officials work with election technology, including the challenges they encounter and solutions they create to overcome them. In the U.S., elections are the product of thousands of local, individualized sociotechnical systems working towards the same goal. This paper is a step towards mapping these intricate organizational structures and recognizing the complex technical work behind running elections. Second, we make recommendations to improve design in election technology to mitigate harm to elections and their workers. Third, our findings support transferable conclusions despite the decentralized nature of election technology used across the U.S. The consistency of election technology challenges across jurisdictions in our sample suggests broader systematic issues with election technology design and deployment that will need policy-focused solutions in addition to building better systems.

\textbf{Roadmap.} We start by providing a background of election work and technology in the U.S. with definitions of key terms used during this paper (\secref{background}). Our related work section covers the history of research on election technology in computer science (\secref{related:testing}) and differentiates our work from preexisting literature  (\secref{related:understanding}). We cover the methods we used to interview participants, protect their confidentiality, and analyze our qualitative data (\secref{methods}). Our findings are organized into three topics based on the major themes from our analysis (\secref{findings}). We first outline the challenges election officials have in leveraging and deploying their election technology (\secref{results:challenges}), followed by documenting the impact those challenges have (\secref{results:impact}), and finish by sharing several ways election officials attempt to overcome these challenges (\secref{results:overcome}). Our discussion focuses on prioritizing needed improvements to election technology based on our findings (\secref{discussion:improving}) and provides direction for future work (\secref{discussion:future}).   

%% file: 2background.tex
\section{Background}

\label{background}

\subsection{Work in U.S. Elections}
\label{background:electionwork}

Work in U.S. elections is decentralized, jurisdiction-specific, and involves the cooperation of several stakeholders.  Capacity among different states can be highly variable due to limited budgets and unique state laws. We broadly explain who these stakeholders are and the responsibilities they have within U.S. elections. 

We use \textbf{election workers} to refer to election officials and poll workers. \textbf{Election officials} are part-time or full-time staff that are paid by the government to oversee election administration. In the U.S., election officials are often referred to local election officials to emphasize their role as members of a smaller, ``local'' jurisdiction within a larger state.  \textbf{Poll workers }are part-time seasonal staff that are typically given flat compensation rates in exchange for several days' of work during an election season. When a voter goes to vote at a polling location in-person, most of the time they are interacting with a poll worker. Poll workers are trained by election officials. Election technology \textbf{vendors} are third-party organizations who develop, make, and sell election technology to jurisdictions.   

An overlooked responsibility for election officials in election work examined in this paper is being accountable for several aspects of election security. Formal threat models of election systems often depend on election officials when vulnerabilities are acknowledged in software to ``detect'' attacks \cite{MITREreport}. Election officials' level of independence over their technology means that insider threats are a nontrivial concern---two criminal cases against election officials providing unauthorized access prove the viability of such an attack \cite{bowerPleaLetterUpdate, sullivanTinaPetersFormer}. Election officials also take on additional roles beyond strict electoral administration depending on the funding status in their jurisdiction, such as managing their own election website or finding their own security training. 

\subsection{Technology in U.S. Elections}
Every U.S. voter’s journey to casting a ballot involves interacting with or depending on technology in some way. Election officials (or poll workers) interface with this technology in tandem. For example, a voter who registers to vote then appears on the internal \textbf{voter registration database} managed by election officials. When that voter arrives to a polling location on election day, they may check-in with an \textbf{electronic pollbook} after a poll worker validates the voter's identification using the same device. Then, a \textbf{ballot-marking device} (BMD), a machine that uses a digital interface to mark a paper ballot, might be used by that voter. That same device would be setup by poll workers earlier that day. Finally, a \textbf{tabulator} would cast that voter's vote, and at the end of the day, a poll worker would retrieve data from this device and shut it down. Any technology run by a poll worker is ultimately overseen by the guidance and support of election officials. 

The combination of election technology used is unique to each state. Several major election technology vendors serve the election community. The U.S. federal government creates standards and tests equipment, but does not purchase the equipment. State or local governments buy the technology from vendors. In rare cases, a state will have the resources to make its own technology, usually the voter registration database.

\subsection{Evolution of U.S. Election Technology} 
\label{background:history}
The Help America Vote Act (HAVA) of 2002 provided funding and established requirements for modernizing voting equipment after issues with equipment during the 2000 U.S. presidential election \cite{iiiWhatHathHAVA, ansolabehereVotingEquipmentBallots}. For example, HAVA required states to have centralized voter registration databases, and accessibility requirements led to the popularity of direct recording electronic voting machines (DREs) \cite{iiiWhatHathHAVA}. The rapid pace of adopting these new technologies immediately brought challenges in operations \cite{neversWhatHappenedHAVA}. It also meant many of these systems had not gone through extensive vetting, prompting computer scientists to begin testing election technology (discussed further in \secref{related:testing}). Partially as a result of this research, most states as of the 2024 election have abandoned DREs in exchange for BMDs or hand-marked paper ballots \cite{verifiedvotingVerifier}. 

These new technologies challenged trust in elections and inspired work on how to verify election outcomes. Security researchers began to think of ways to integrate end-to-end-verifiable protocols into election technology for voters to prove their vote has been counted \cite{benalohElectionGuardCryptographicToolkit2024}. Risk-limiting audits (RLAs) were introduced to provide a bound on statistical risk of an audit of ballots failing to catch a wrong outcome as reported by tabulators \cite{lindemanGentleIntroductionRiskLimiting2012}. These developments are notable in the evolution of election technology because current solutions require additional labor from election officials or educating voters to be successful. RLAs require ``rigorous ballot accounting'' \cite{WhatRiskLimitingAudita}, while prior work has shown voters rarely verify votes when unguided \cite{bernhardCanVotersDetect, hiltWhyJohnnyChecks2026} and that stakeholders question the practical utility of current verifiable protocols \cite{haneyUSElectionExpert2026}.

%% file: 3relatedwork.tex
\section{Related Work}

\subsection{Testing Election Technology}
\label{related:testing}
In usability studies, research on election technology predominantly focuses on measuring and improving \textit{voter} experiences. As ballot-casting election technology like ballot-marking devices (BMDs) and direct recording electronic (DRE) systems became more popular in the early 2000s in the U.S., there was growing concern that poorly designed systems introduced at scale could disenfranchise voter intentions through mistakes or confusion \cite{kingrothDisenfranchisedDesignVoting1998, conradElectronicVotingEliminates2009, herrnsonImportanceUsabilityTesting}. Researchers began to discuss and analyze the usability of voting technologies \cite{bedersonElectronicVotingSystem2003, robertsonVotercenteredDesignVoter2005, selkerVotingUserExperience2003, olemboEVotingSystemUsability2013, acemyanAssessingUsabilityHart2017a}. Accessible  voting technologies were developed as  alternatives to paper ballots for voters with certain disabilities \cite{crossPrimeIIIUser2007, gilbertAccessibleVotingOne2011, leeUniversalDesignBallot2016}. Recent work continues to improve and expand upon prior research, thinking about other aspects of voter usability like verification \cite{hiltWhyJohnnyChecks2026, bernhardCanVotersDetect} and internet voting \cite{markyImprovingUsabilityUX2020}. Overall, there is room to explore how election \textit{workers} navigate election technology and identify the usability challenges they face.

Security research on election technology has largely focused on the casting and tallying process (e.g., BMDs and DREs)  \cite{specterSecurityAnalysisDemocracy2021, feldmanSecurityAnalysisDiebold2007, specterBallotBustedBlockchain2020, avivSecurityEvaluationESS2008, kohnoAnalysisElectronicVoting2004, debantReversingBreakingFixing2023, hainesVestigialVulnerabilitiesDeployed2025} and constructing new, secure voting systems for verification and validity \cite{adidaHeliosWebbasedOpenAudit, benalohElectionGuardCryptographicToolkit2024, clarksonCivitasSecureVoting2008}. These works are often very heavy in cryptographic theory and usability for election officials is typically not considered. Additionally, these studies tend to focus on only one part of the election technology ecosystem at a time, overlooking the interactions between the entire system. Election officials oversee many technologies while administering elections, each of which could have vulnerabilities. Voter registration databases, electronic poll books, reporting systems, and even internal office software can be hacked and violated. Second, formal threat models in election security have listed humans as the primary mechanism to catch certain attacks or mistakes \cite{MITREreport}. Isolating the technology from the sociotechnical system and the factors that strain it (e.g., hate and harassment) does not capture security holistically.

\subsection{Understanding Election Work}
\label{related:understanding}

Political scientists have conducted substantial research on the work of election officials and poll workers. A significant amount of research in this area seeks to understand problems in election administration by using theories in political science. Alvarez and Hall argued that principal-agent theory could explain issues election officials have managing poll workers \cite{alvarezControllingDemocracyPrincipal2006}. Montjoy and Slaton claim that election work is interdependent and explain how this interdependence presents an ethical problem for accountability \cite{montjoyInterdependenceEthicsElection2002}. Another approach political scientists take is using quantitative methods to compare worker practices with election outcomes, like Spencer and Markovits using observational data to understand how lines form at polling locations \cite{spencerLongLinesPolling2010} or Atkeson et al. exploring poll worker discretion in identification \cite{atkesonWhoAsksVoter2014}. Crucially, this research often overlooks the relationship between election work and election technology or the broader sociotechnical systems behind election work.

Concerns about elections have provided an incentive for organizations and academics to conduct longitudinal empirical measurements of election work. In the U.S., there are several general surveys on election work that take place every year or after every election. The U.S. Election Assistance Commission (EAC) has election officials fill out the annual Election Administration and Voting Survey (EAVS) \cite{eac2025eavs}. Since 2018, the Elections \& Voting Information Center (EVIC) conducts a yearly survey sampling local election officials \cite{gronkeDecliningJobSatisfaction2025}. These surveys surface challenges election officials face---like funding crises, labor shortages, and low salaries---but ask few questions about election technology. In 2021, The Brennan Center for Justice began surveying local election officials and included questions about digital threats \cite{brennancenterLocalElectionOfficials2025}. In 2025, roughly one-third of election officials they surveyed reported experiencing harassment \cite{brennancenterLocalElectionOfficials2025}. Of those election officials, 59\% experienced threats over the phone, 39\% through social media, and 30\% through email \cite{brennancenterLocalElectionOfficials2025}. Due to the quantitative nature of the survey, it is not clear how election officials react to these digital threats, or whether an increase in digital threats has changed how election officials engage with election technology. 

Despite the importance of understanding challenges with election technology in election work, there is minimal research on this subject in HCI, CSCW, and other human-centered computer science disciplines. In 2015, Boulus-Rødje and Bjørn interviewed election officials in Denmark to understand election organization through the framework of ``nomadic knowledge'' \cite{boulus-rodjeDesignChallengesSupporting2015}. Their findings described what technologies these election officials used to share information about procedures (e.g., email or text) and focused on the distribution of information in election work. Scala et al. developed and tested a training module for security threats for U.S. poll workers, demonstrating an increase of awareness after completing the module, but this work did not extend to election technology used by election officials \cite{scalaPREPARINGPOLLWORKERS2023a}. Panahai et al. interviewed ``participants familiar with'' biometric voter verification in Afghanistan to understand challenges with implementation \cite{panahiTheyHaveOverlooked}. While the primary goal of this paper was to understand how biometric voter verification impacted Afghan elections, not election work, one finding did include the mention of usability issues leading to poll worker mistakes \cite{panahiTheyHaveOverlooked}. There is a gap in understanding the interactions between election workers and election technology within these sociotechnical systems.

\subsection{Technology Use in Public-Facing Work}
 \label{related:public}
Security, safety, and privacy are a concern when technology and public-facing work meet. Often, people doing public-facing work, like journalists, activists, and civic employees, operate with limited funding. The unique combination of risk in public-facing work puts stress on sociotechnical systems and impacts how workers in these spaces leverage technology. Research in HCI and CSCW has repeatedly affirmed the need to dissect and document how these different types of risk lead to novel challenges. 

Researchers have found that concerns about safety and privacy impact how people use digital communication channels in public-facing work. Moran et al. documented how a research team's affiliation with election work meant that their communications could be discoverable by FOIA requests as well \cite{moranPrivacyTransparencyNavigating2025}. They found that collaborative work environments are strained when informal digital communication channels can become public record \cite{moranPrivacyTransparencyNavigating2025}. McGregor et al. found journalists alter their practices with digital communication based on the risk of leaks or hacking and have to handle privacy concerns when talking with sources \cite{mcgregorWouldYouSlack2017}. Tadic et al. found that activists with low levels of protection self-censored their online communication as a form of security \cite{tadicICTUseProminent2016a}.

Prior work has also sought to understand what prevents public-facing workers from using privacy and security features.  Consolvo et al. found that transience and limited budgets prevent political campaign staff from adopting privacy-enhancing tools and standardized technology practices \cite{consolvoWhyWouldntSomeone2021}. McGregor et al. found that usability of digital security features inhibit journalists from using them \cite{mcgregor_investigating_2015}. These works support the need to identify unique challenges in leveraging technology for public-facing roles that democracies depend on.  

%% file: 4methods.tex
\section{Methods}
\label{methods}
We interviewed (N=14) and surveyed (N=45) U.S. \leo{s} to understand the technical challenges they face in conducting U.S. elections. In this section, we explain the limitations of this study, recruitment processes, interview and survey procedures, and data analysis. We cover the steps we took to maintain our participants' confidentiality, including obtaining a NIH Certificate of Confidentiality, and the preparation we took before beginning our data collection, including observing real trainings and engaging with an experienced \leo. In the Appendix, we provide aggregated statistics of survey results with the exact question presented to participants along with the semi-structured interview outline for reproducibility of charts and replicability of our analysis.

\subsection{Preparation and Recruiting}
\label{subsec:recruit}
Our preparatory work began with all authors attending a poll worker training session led by an election director in 2024. Two members of our research team participated as poll watchers in the 2024 U.S. presidential election to further learn more about procedures during an active election. Before starting any data collection, we worked with an expert with a background in election administration to inform our research goals. In early 2025, we conducted a pilot interview with this expert and identified initial themes from that data to identify problem areas of focus. We collected feedback on our analysis from this expert to solidify our interview outline and research questions.

For interview and survey participants, our recruitment materials advertised this study as an exploration of how election officials interact with their technology. We recruited through multiple methods to increase our coverage among different types of jurisdictions (e.g., more or less populous, political lean, etc.). Our process started through outreach within our community and professional connections, which then led to participant-driven, or snowball, sampling. To avoid some of the limitations of snowball sampling, we asked participants in each state if they could put share our recruitment materials with \leo{s} who are less likely to be represented (e.g., \leo{s} in smaller or economically disadvantaged jurisdictions). We then moved to recruiting through professional email lists for all 45 survey participants and provided an optional interest form for an interview at the end of our survey described below (\secref{subsec:survey}). Nine of the 14 interview participants were recruited using the survey or via advertisement of the survey and five through personal contact. In total, 50 unique election officials were interviewed or surveyed. No financial incentive was provided to participate in this study to comply with certain U.S. states' laws restricting when government employees can receive compensation and from whom.

\subsection{Participants}

We interviewed 14 \leo{s} that represent six states in the U.S. There are 2 state-level \leo{s} and 12 local \leo{s}. Most participants are the director for their jurisdiction (8, including 2 municipal directors who are in charge of a city within a larger county), followed by deputy directors (4) who are defined as being adjacent to second-in-command. For population, we binned 2023 U.S. Census Bureau population data \cite{ers_rucc2023} into ranges used on other U.S. Census Bureau resources \cite{elnasser2017counties}. We do not provide the names of the states to lower the risk of identification. However, we do provide the U.S. Census regions \cite{census_geo_terms} represented by our interview participants along with other information relevant to the study (Table \ref{table:interviewdata}). We note that our interview participants come from  politically significant states that represent many U.S. residents. The median number of electoral votes in the states that our participants represented was greater than 15. The national median is roughly 8 electoral votes per state.  

Our online survey participants represent 20 states and the average number of participants per state was 2.36 (median of 1; range of 1 to 9). Like with our interview participants, we do not provide the names of states but do share relevant information that has a lower risk of identification (Table \ref{table:surveydata}). 

\input{tables/qual-participants}
\input{tables/quant-participants}

\subsection{Interview Procedure}
We conducted interviews from January 2025 to September 2025. Interviews were conducted over Zoom, which allows end-to-end encrypted recording locally. One person, the first author, conducted all the interviews. The interviews ranged from 44 to 85 minutes ($\mu$=53). Each interview started by reminding the participants of their data rights during and after the interview and asking if they had any questions. Our interviews were designed to cover our research questions, adapting to the experiences of the participant in a semi-structured manner. After each interview, the audio was locally transcribed using an open-source LLM. The transcriptions were checked against the interview recordings to correct any errors caused by the LLM.

\subsection{Survey Procedure}
\label{subsec:survey}
During the interview process, we developed a survey to sample more \leo{s} and validate the transferability of our interview findings to a larger number of states. The survey took approximately 20 minutes and included yes-or-no, Likert scale, and open-ended questions.

Seventy-six participants attempted the survey, starting with an attestation asking if they had ever been an \leo. We filtered these 76 participants to 45 based on whether they had clicked through all the questions using a feature on our survey platform. These records had answered the majority of questions (the 25th- to 75th-percentile of skipped questions was 0 to 3, or 0\% to 5\%). We gave participants the option to skip questions if they were not comfortable sharing certain information, leading to numbers that do not add up to 45 in some cases.

\subsection{Analysis}
We used qualitative analysis to examine our interview and survey data together. Election administration is unique for each jurisdiction. \Leo{s} have different election technology and procedures in each state. For this reason, we needed an interpretive approach to capture broader conclusions from our data. As a result, we used deductive coding aligned with reflexive thematic analysis \cite{braunclarke2019reflecting}. 

After each interview, the first author would review transcript data and take notes. After five interviews, major ideas started to repeat and the first author developed a preliminary codebook along with initial themes. These themes were reviewed with the research team and a survey was designed with the first, second, and last author to examine the transferability of these themes to a larger number of states. After completing the interviews and the survey, the first author continued to code transcripts. Once all interviews were coded, these codes were integrated into the initial themes and themes were combined or added as necessary. The survey data, already organized based on the initial themes it was built upon, supplemented this final list of themes. Our analysis presents these themes organized into 17 groups. 

\subsection{Ethical Considerations}
Using the framework proposed by Warford et al. \cite{warfordSoKFrameworkUnifying2022}, we classify our participants as at-risk due to their sensitive position within the U.S. democratic process. We were especially cautious of the threat of doxxing due to the political violence experienced by \leo{s} across the U.S. We outline how we protected our participants' confidentiality and safety during data collection and analysis. 

Our study design was approved by our Institutional Review Board (IRB). We waived signed consent to prevent documentation that might trace a participant back to our study. We were particularly mindful about the potential harm of public records requests (e.g., FOIA) on this population and affiliated researchers, as documented by Moran et al. \cite{moranPrivacyTransparencyNavigating2025}, and therefore obtained a NIH Certificate of Confidentiality \cite{nih_coc_intro} to add an additional layer of protection from certain types of legal requests. We used end-to-end encrypted platforms when possible, including storing our interview audio on KeyBase and recording interviews locally on Zoom. The tools used to transcribe and code interviews processed data locally.

In presenting demographic data, we aggregate our survey responses and separate demographic data from the responses. For our interview participants, we provide regions instead of states to make sure no combination of region, state government, and population is traceable to a particular state. We recognize that participant aliases (e.g., ``\textit{P00 said}'') help support that our qualitative analysis is representative of the entire sample. In instances where we felt a provided anecdote is critical to our analysis but unique enough to be associated with a specific state, or could risk a participant's safety, we took a hybrid approach to selectively omit aliases (e.g., ``\textit{a participant said}'') and abstract details as seen in other work with politically sensitive subjects \cite{consolvoWhyWouldntSomeone2021} out of an abundance of caution. 

\subsection{Limitations}
This work has the standard limitations shared by all interview and survey studies. We are not capturing the views of \leo{s} who might be dissuaded from participating due to the politically sensitive nature of our topic.  We made targeted efforts to build trust and interview \leo{s} who are historically underrepresented and have fewer resources, but the representation of our samples are not comprehensive. For example, our interview sample does not include anyone from the Midwest. Our survey sample is missing input from 30 out of 50 states. We try to offset this bias by emphasizing that the states in our sample represent a majority of registered voters in the U.S.  

While we try to incorporate geographic information about the density and wealth of our jurisdictions, we do not extend this analysis to examine relationships between problems with election technology and other sociodemographic variables like poverty or race that might have implications on civil rights. We do not include variables like gender or race of \leo{s} that might lead to different outcomes on seeking technical help and support. This is primarily because our sample size would not support analyses this granular. Future work should investigate the relationships between sociodemographic factors and challenges with election technology.

%% file: tables/qual-participants.tex
\begin{table*}[ht]
\caption{Information about the local jurisdictions represented by our 14 interview participants. NA indicates data that does not apply to a state-level \leo.}
\begin{center}
\begin{tabular}{llllllll}
\toprule
 \textbf{ID} & \textbf{Census Region} & \textbf{Position}           & \textbf{\# of Staff} & \textbf{Population} & \textbf{State or Local} & \textbf{State Government} \\
\midrule
\rowcolor[HTML]{EFEFEF}
P01         & South             & Director           & 6 to 10                   & 100-500K         & Local          & Republican                                        \\
P02         & South             & Deputy Director    & 3 to 5                     & 20-100K          & Local          & Republican                                        \\
\rowcolor[HTML]{EFEFEF}
P03         & South             & Director           & 2                    & 5-20K            & Local          & Republican                                        \\
P04         & South             & Director           & 2                    & 5-20K            & Local          & Republican                                        \\
\rowcolor[HTML]{EFEFEF}
P05         & South             & Deputy Director    & 3 to 5                     & 20-100K          & Local          & Republican                                        \\
P06         & Northeast         & Director           & 1                     & 20-100K          & Local          & Democratic                                        \\
\rowcolor[HTML]{EFEFEF}
P07         & South             & Director           & 6 to 10                    & 100-500K         & Local          & Republican                                        \\
P08         & Northeast         & Municipal Director & 3 to 5                     & 500K+            & Local          & Democratic                                        \\
\rowcolor[HTML]{EFEFEF}
P09         & Northeast         & Deputy Director    & NA                   & NA               & State          & Democratic                                        \\
P10         & Northeast         & Municipal Director & 3 to 5                     & 500K+            & Local          & Democratic                                        \\
\rowcolor[HTML]{EFEFEF}
P11         & West              & Specialist         & NA                   & NA               & State          & Democratic                                        \\
P12         & West              & Assistant Clerk    & 3 to 5                      & \textless{}5K    & Local          & Democratic                                        \\
\rowcolor[HTML]{EFEFEF}
P13         & West              & Deputy Director    & 10+                      & 500K+            & Local          & Democratic                                        \\
P14         & Northeast         & Director           & 1                      & 20-100K          & Local          & Democratic                                       \\
\bottomrule
\end{tabular}
\end{center}
\label{table:interviewdata}
\end{table*}

%% file: tables/quant-participants.tex
\begin{table}[]
\caption{Demographic summary of the survey participants. \textit{Local area} is a self-reported description of their jurisdiction. NA indicates data that does not apply to a state-level election worker.}
\begin{center}
\begin{tabular}{@{}llll@{}}
\toprule
\textbf{Demographic}   &                    & \textit{n}  & \%   \\
\midrule
\textbf{Local Area}    & Rural              & 13 & 28.9 \\
              & Suburban           & 17 & 37.8 \\
              & Metropolitan       & 13 & 28.9 \\
              & NA                 & 2  & 4.44 \\
\textbf{Census Region} & South              & 14 & 31.1 \\
              & Midwest            & 10 & 22.2 \\
              & Northeast          & 11 & 24.4 \\
              & West               & 10 & 22.2 \\
\textbf{Experience}    & \textless{}5 years & 11 & 24.4 \\
              & 5 to 10            & 10 & 22.2 \\
              & 10+                & 24 & 53.3 \\
\bottomrule              
\end{tabular}
\end{center}
\label{table:surveydata}
\end{table}

%% file: 5findings.tex
\section{Findings}
\label{findings}

\subsection{Challenges with Leveraging and Deploying Election Technology}
\label{results:challenges}

\subsubsection{\textbf{Hard-to-use election technology interfaces}} 
\label{results:sub:design}
The usability of election technology interfaces commonly challenges election officials in their offices and during election day. Most interview participants gave examples of how their election technology is confusing or hard to use in a way that impacts their operations. We organize these challenges by those experienced by election officials and those typically faced by poll workers. 

For election officials, interacting with voter registration databases came up repeatedly in interviews as a source of frustration. Broadly, voter registration database design challenges focused on slow or inefficient software causing input errors, confusing voter statuses or reports, and redundant or duplicate processes. S13 said about their voter registration database in an open-response question about what election technologies they wish were easier to use, ``[the voter registration database] \textit{can be clunky to use, and if you are not sure how to search properly, errors can occur}.'' For one election official, a search drop-down menu that is too slow and alphabetized instead of best-matched has led to election officials clicking on the wrong address for voting records. For another, lack of data validation led to situations like typing date of birth into a last name field, and being able to press submit. A third election official described how a lack of labeling can lead to challenges in voter record management: 

\begin{quote}
    ``We have all these duplicate records, where one is active and the other is cancelled. It's kind of cluttering, and it's very easy for a new person dealing with registration to mistakenly reactivate the cancelled duplicate record as a result... they're stacked one on top of each other and there's no sorting based on the status.'' ---\textit{P14}
\end{quote}

Election technology that is hard to use also impacts poll workers at polling locations. Problems with electronic pollbooks, scanners, and printers center around the interfaces and setups of these devices overwhelming or confusing poll workers. S14 said in an open-response question that the hardest election technology to train for was their scanner ``\textit{simply because of bad UI} [user interface].'' Another survey respondent, S23, shared that some locations do not set up ADA machines because not everyone understands how to use them. Another election official explained why they want better designed polling location software:
\begin{quote}
    ``I would love for interfaces that less technologically inclined poll workers could interact with... the actual essentials and not give them information overload with all the advanced options right there to make them afraid that they'll mess up and hit the wrong button.'' ---\textit{P04}
\end{quote}

Across poll workers and election officials, the challenges described to us originate from a user of election technology being allowed to make the wrong choice on an interface. Sometimes, the problem is how selections are presented---``\textit{they thought they closed the polls, but instead, they just powered off the device because the} [touchscreen] \textit{buttons were next to each other},'' shared P04. However, a bigger concern among participants was when users make an intentional choice that is wrong by accident without any confirmation or safety mechanism to prevent it. The reason for concern is that then election officials have to catch the mistake. One participant's voter registration system allows voters to be registered in certain edge cases when registration deadlines have passed. ``\textit{We might not catch it... we all make mistakes, so, the likelihood of us making that mistake is likely}.'' Another system creates a duplicate record when the poll workers make the wrong choice at check-in, an event so common that they ``\textit{double-check}'' for it internally on a daily basis. One participant found out about a mistake that is now ``\textit{included in} [their] \textit{instructions}'' when an experienced poll manager was given an option to hit ``\textit{cancel}'' or ``\textit{proceed}'' on an electronic poll book after having technical difficulty. ``\textit{This manager just hit cancel}'' to let the voter vote. ``\textit{When} [the poll workers] \textit{came back to do their numbers, they were off by one},'' meaning that vote record was potentially lost. This example highlights how even experienced poll workers can be confused by the options given on current election technology when they see a new problem or error.

\subsubsection{\textbf{Election technology not working reliably}} 
\label{results:sub:notworking}
Election technology can be used thousands, if not tens of thousands, of times on election days. Election officials emphasized the need for durability and reliability when they are making a costly investment in these devices. However, our participants shared various ways in which the election technology they deploy can be unreliable when it is needed most. In an open-response question on our survey, S02 shared the hardest part of training others on election technology is ``\textit{a healthy skepticism about whether the equipment is actually working properly.}'' P05 said of the electronic pollbooks, ``\textit{you've got to hold your tongue right to make things work... it's just finicky}.'' P05 also shared of an election day when a ``\textit{memory card just went defective like during the middle of the day}'' that caused stress because it was only one of two scanners they had at the polling location.

Often, election technology stopped working reliably when systems were pushed to their limit. P11 said the search function for checking voters in at polling locations is ``\textit{very process heavy}.'' Recently, a  campaign urged more than the usual number of voters to show up on election day. ``\textit{When a lot of counties are doing the same thing, it really bogged down the system},'' causing delays and wait times. Multiple interview and survey participants noted syncing issues with systems like electronic pollbooks that sometimes connect to voter registration databases to provide real-time information on election days. This problem is worse in rural regions where cellular and internet connection is tested the most. ``\textit{We have challenges in some of the rural parts of the state... there's constantly issues with connectivity and getting those} [election night] \textit{results},'' P09 said.

\subsubsection{\textbf{Poll workers seek help for technical problems}}
\label{results:sub:support}

Election officials expect poll workers and less experienced election officials to need assistance during elections. In our survey, 44\% of election officials did not feel confident about poll workers' ability to troubleshoot election technology. S34 shared that the hardest part about training was troubleshooting---``\textit{every situation is different, so it's hard to train on, `if this happens.'}'' Election officials anticipate needing to provide support because of the limitations on training. When talking about mistakes poll workers make, P07 shared, ``\textit{you can't fault them because they only work, you know, once or twice a year}.'' Upgrades only exacerbate this problem---in our survey, 55\% of election officials agreed that new or updated election technology takes a while to understand.         

Poll workers were most likely to struggle with the ``\textit{little things}'' like turning devices on or off, connecting devices to Wi-Fi or Bluetooth, and navigating interfaces on the electronic pollbooks and scanners. While these may seem like simple tasks to someone with more technical skill, election officials stressed that not everyone has the same experience. Assumptions about poll workers' baseline familiarity with technology are integrated within these systems---assumptions that are sometimes wrong. P01 shared how their electronic pollbooks are easy to use ``\textit{if you have a base level of technological expertise...like you've used an iPad before}.'' Some poll workers ``\textit{struggle with that just because they haven't} [used an iPad before].'' 

\subsubsection{\textbf{Many poll workers have unique accessibility needs}}
\label{results:sub:senior} 
Over half of all poll workers in the 2024 election were over the age of 61, and nearly 30\% over the age of 71 \cite{eac2025eavs}. Some regions have older populations than others. ``\textit{My poll workers probably start 55 years of age and go up},'' said P03, who is in a rural jurisdiction. Since 2020, poll workers have become older on average \cite{eac2025eavs}. This fact was observed by one survey participant: 
\begin{quote}
    ``Our poll workers are getting older and less technology capable. As technology evolves, it can become more difficult for poll workers to keep up on the changes.'' ---\textit{S13}
\end{quote}
When survey participants were asked what the hardest part of training others to use election technology was, several mentioned the age of the poll workers. ``\textit{It is hard to teach 70-and-80-year-olds how to operate a computer for the first time},'' mentioned S12. The expectation of older poll workers is embedded into training. ``\textit{You also have to deal with the nuances of who are your poll workers... here your average poll worker is 70 and 80 years old},'' shared P02. 

Notably, the challenge with senior poll workers using election technology is only loosely related to digital fluency. Some problems had to do with physical accessibility.  P03, mentioned earlier as a participant in a rural jurisdiction with many senior poll workers, described how the expectations of working ``\textit{fourteen, fifteen hours}'' on election day takes a larger toll on their older poll workers. Senior poll workers have a harder time completing tasks that require vision and dexterity, meaning they struggle to find ports or switches that ``\textit{charge and turn on}'' technology, as P10 and others explained.

\subsubsection{\textbf{Lack of professional technical support}} 
\label{results:sub:ITsupport}
Election officials from multiple states mentioned a systematic lack of technical support at the local level, especially for rural regions. P03 said, ``\textit{my county does not have an IT department... but half of} [other counties] \textit{are my size without IT departments.}'' P04 mentioned, ``\textit{I am the closest thing my county has to an IT department},'' and P06 said that their IT help desk was ``\textit{affectionately}'' called the ``\textit{lack of help desk}.'' One of our state-level interview participants provided a broader look at the availability of IT support at the local level:
\begin{quote}
    ``So we've got cities and towns that, you know, don't have an IT person. They've got maybe a third-party vendor they work with, and maybe the tax collector or the recreational director knows computers pretty well, and so they kind of act as a de facto IT person. But we know that the resources at the local level are very, very scarce.'' ---\textit{P09}
\end{quote} 
In the absence of support, election officials are faced with the challenge of troubleshooting their own technical problems and making choices about the technology in their offices. This creates a need for technical skill within the role---even when these requirements are not advertised or compensated as such. ``\textit{Hopefully she's computer literate, or I'll be real honest, I probably won't hire her},'' said P03, who does not have an IT department, on the skills they look for in hiring an election official.

\subsubsection{\textbf{Dependency on the election technology vendor market}}
\label{results:sub:vendors}
Aside from a few exceptions, states in the U.S. do not have enough resources to develop their own election technology to fit the exact needs of their jurisdictions. Most election officials rely on private vendors to sell them (or their states) the technology they need to conduct elections. This dependency creates challenges when critical vendor support is hard to get, a lack of uniformity across vendors leaves some election officials much worse off than others, and a limited market with little competition leads to stifled innovation. 

P09 emphasized, ``\textit{working with the vendors is really, really important, especially if we need changes to systems}.'' However, several interview participants shared experiences which it was hard to get proper support from vendors when it was needed. P06 said a vendor canceled a project after working on it for years and their state had to ``\textit{start the process all over again}.''  P12 shared a time when they waited days to hear back from a vendor after encountering a time-sensitive error. They added, ``\textit{if I want good advice, I have to talk to other counties. I can't actually talk to the vendor for the best advice, which is really frustrating.}'' How election officials rely on their  communities will be explored more in (\secref{results:overcome}). Some election officials said they have had vendors who charge for updates, making it hard to receive support when budgets are low or cut. ``\textit{The biggest complaint with the old system is that the vendor could change anything we wanted, but they would charge us money},'' said P08. 

Based on positive and negative feedback about vendors we received from participants, the market has a lot of variability in quality that can lead to some states being better equipped than others. Given that many states choose the election technology for their election officials, this led to frustration over vendor choice at times. P01 shared they want to be part of the next selection of election technology for their state because ``\textit{some are a lot more usable than others, and they've all solved problems very, very differently.}''

Election technology is a limited market due to strict regulations, a high barrier to entry, and low profit potential. Among election officials who spoke of this problem, there was a desire to ``\textit{develop a market that has maybe more incentive for more people to jump in the game},'' as P13 put it. P04 said the lack of competition ``\textit{stifles innovation}'' and that they ``\textit{do think it is harmful to election administration and election integrity as a whole for how few vendors are currently in the space}.''

\subsubsection{\textbf{Standardized election technology has trouble adapting to local law}} 
\label{results:sub:standard}
A consequence of (\secref{results:sub:vendors}) is that election officials sometimes need to find workarounds to adapt their election technology to the laws of their local jurisdiction. For example, when technology used at the state level needs to work for special or local elections, local election officials do not have the ability to modify their systems to meet their requirements. An automated registration check did not account for a local election deadline in a participant's jurisdiction. They mentioned how their system allows them ``\textit{to continue to register people for this election when we're not supposed to}'' and that they have to train the staff to manually check for this edge case. Fixing these edge cases is hard, and usually requires a lot of resources and a dedicated vendor privy to the needs of that specific jurisdiction: 

\begin{quote}
    ``Building a system that does all the things that we need it to do based on the laws, but without losing any of the functionality we have, and actually gaining some, that seems to be the real tricky part of the build.'' ---\textit{P06}
\end{quote}

\subsubsection{\textbf{Technology solutions are limited by election security and integrity needs}} 
Election officials and poll workers are often tasked with extensive manual data entry and processing. Some of these processes are repetitive, for example, forms that require counting and cross-checking numbers on multiple voting machines, or inputting voter data manually when the data exists elsewhere. Additionally, workers get stressed about them. ``\textit{Math is something that scares a lot of people},'' explained P06. These tasks exist for a reason---due diligence is an important part of ensuring voting rights and maintaining election integrity. Tracing decisions back to humans allows for accountability. However, election officials shared that some of these tasks could be automated while maintaining oversight. 

The reason these tasks are not automated has less to do with their feasibility than whether voters or election officials themselves could trust these systems. P02 described a process where poll workers have to fill out a form for each piece of equipment and then compile that form into a ``\textit{master form}.'' The data on these forms are present in other forms that are also filled out. P02 said processes like these could definitely be streamlined, but: 

\begin{quote}
    ``I don't see how it can ever get streamlined without automation. I don't see how the voters will ever accept that automation. There are certain things that are just going to have to be manual because the voters are not going to trust a machine to do it, you know?'' ---\textit{P02}
\end{quote}

\subsubsection{\textbf{Policy works against technology needs}}
\label{results:sub:policy}
Election officials are often subject to the decisions of stakeholders in their local jurisdictions and state lawmakers. As our participants explained, these stakeholders are not always aware of or able to understand the complex relationship between various election technologies and the needs of election officials. 

At the state level, the disconnect between lawmakers and election officials is exacerbated when politics play a role. When a state legislature and governor clashed, the consequence was ``\textit{holdups with receiving funding on upgrading and even modernizing our voter registration system}'' for one participant. Another participant explained that when the state changed the laws for how to handle an at-risk population in the voter registration database, that ``\textit{we had the lawyers review our policy and made a policy because you know, registration} [for this group] \textit{doesn't stop just because nobody's given us instructions}.'' State legislatures are farther removed from the needs of local officials, and can make choices that go directly against improving election officials' use of technology:
\begin{quote}
    ``I just sometimes wish that the legislature understood what they were voting on and what the implications are, because between, you know, the laws being kind of willy nilly, and then the technology not matching up, it just creates kind of a perfect storm for a system that is less than user friendly.'' ---\textit{P06}
\end{quote}

\subsubsection{\textbf{Limited funding and the loss of institutional resources}} 
\label{results:sub:funding}
Funding challenges election technology use in three ways. First, there is a baseline lack of funding to support election technology at the local, state, and national level. Second, this lack of funding has been exacerbated by recent cuts to federal programs designed to support election technology. The lack of federal assistance combined with uncertainty in local budgets creates a third challenge when election officials have to invest in expensive technology for long-term use. 

Locally, decisions like approving budgets have to go through city or county officials who might be dealing with separate needs and limited funding (e.g., fire stations and community projects). One participant shared how they ``\textit{know our county is in a really bad economic situation}.'' P07 said that technology upgrades ``\textit{still have to get my} [county] \textit{board's approval... because it involves county dollars}.'' S45 mentioned that a new version of a scanner has better error messages but they are ``\textit{not sure my county can afford to upgrade}.'' Elections, while essential, might not be as pressing as immediate local needs for government services. And because most states decide the election technology used state-wide, when states lack funding, that impacts local election officials, too. P14 described a feature in their legacy system that was not implemented in the new version of their voter registration database. ``\textit{They ran out of money... so I think it's probably not until the next version} [of this software] \textit{that they will address} [this feature].'' 

Receiving national funds to upgrade election technology is also a struggle for local election officials. P14  mentioned that ``[U.S.] \textit{Congress will appropriate money and it will go to the states, and then sometimes it gets passed down} [to local communities], \textit{sometimes it doesn't}.'' Recent funding cuts \cite{samara_angel_trumps_2025, cassidy_us_2025} to institutions like the Cybersecurity and Infrastructure Security Agency (CISA) and programs like the Multi-State Information Sharing and Analysis Center (MS-ISAC) mean election officials have fewer people to consult when looking for technology support. Election officials had gone to CISA for reassurance about the safety and security of their systems, when buying new equipment and software, and as a general resource for many questions related to technology, even those not directly related to cybersecurity.  

The loss or limitation of funding at all levels had a strong impact on election officials trying to plan future election technology purchases. Election technology is a large investment for these governments, and election officials need purchases to last decades. States save for years to upgrade election technology and usually only have enough funds right before the technology becomes obsolete. P09 shared their state works ``\textit{continuously}'' to replace electronic poll books that have a five-to-six year lifespan. P08 said they wish their state did not buy a new model of election technology ``\textit{two years ago right before this new model came out because the new model is a lot better... I know we're stuck with the old one for a while}.'' P06 spoke about the pressure to modernize their election technology by external stakeholders, and said ``\textit{anytime we're talking about technology and improvements... part of the conversation is just like, where does that money come from, and the manpower, in order to achieve those things}?''

\subsection{The Impact of Election Technology Challenges on U.S. Election Officials}
\label{results:impact}

\begin{figure}
    \includegraphics[width=1\linewidth]{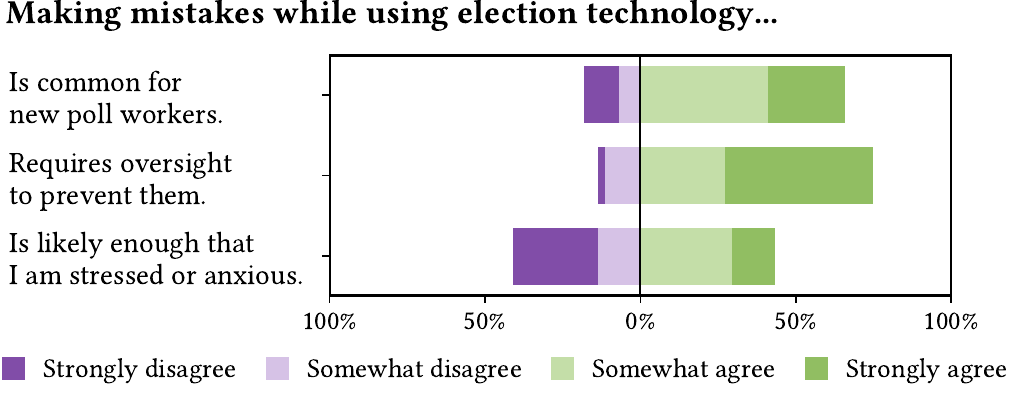}
    \caption{Likert-scale results of the 45 election officials who participated in the online survey. The left side of the chart shows disagreement and the right side shows agreement.}
    \label{fig:making_mistakes}
\end{figure}

\begin{figure}
    \includegraphics[width=1\linewidth]{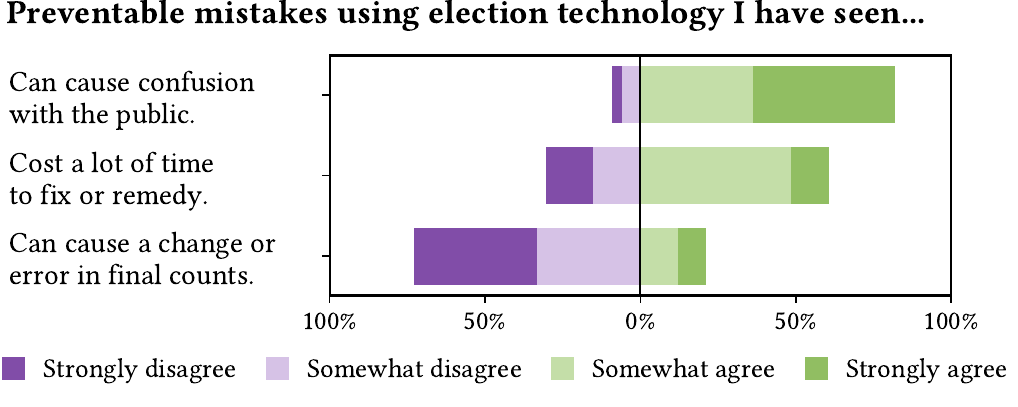}
    \caption{Likert-scale results of the 33 election officials who participated in the online survey and answered ``yes'' to seeing preventable mistakes while using election technology (73\%). The left side of the chart shows disagreement and the right side shows agreement.}
    \label{fig:fixingmistakes}
\end{figure}

\subsubsection{\textbf{Preventable mistakes happen when using election technology}}
\label{results:sub:mistakes} In our findings about design, accessibility, and resources in (\secref{results:challenges}), we previewed the broader impact these challenges have on elections---they create an environment prone to users making mistakes. Thirty-three (roughly 73\%) of our survey participants answered ``\textit{yes}'' to witnessing preventable mistakes made while using election technology. A majority of our survey respondents agreed that making mistakes using election technology requires oversight to prevent them, and nearly 50\% agreed that their likelihood is high enough to make them stressed or anxious during election seasons (Fig. \ref{fig:making_mistakes}).

Even when usability is not an issue, complex election technology can still overwhelm election officials in high-stakes or fast-paced situations, like on election day or during tabulation. Election officials shared that they and their poll workers have to make a lot of decisions when selecting options on these systems. When thousands of votes are coming in or there is a line out the door, the added stress and time pressure can lead to mistakes. P11 shared that ``\textit{more of the errors happen when receiving ballots, we have a lot of statuses we're trying to track... when officials are receiving a lot of ballots, that's a difficult time to pay attention to details like that, you know?}''

\subsubsection{\textbf{Election officials have difficulty keeping software on systems up-to-date}}

Difficulty updating election technology software is another consequence of the design and accessibility challenges election officials face. Software updates are a major responsibility for election officials, and  critical updates can affect election outcomes \cite{staff_voting_2023}. Our participants shared that the extensive amount of skilled labor required to update voting technology is a lesser-known obstacle to election security. Some participants explained how these updates can take a ``\textit{very long time}'' (P01) because they sometimes depend on election officials, who are not required to have advanced technical backgrounds, to administer the software updates manually. 

\begin{quote}
    ``This upgrade process was not simple... There were a lot of assumptions made when we got our new system that [local officials] knew how to secure it because we secured the previous voting system. I'm comfortable doing [those updates] but can you imagine some [other election officials] trying to? I mean, you're gonna break the devices.'' ---\textit{P01}
\end{quote}

Similarly, minor updates or changes to interfaces, especially for poll workers who do not have technical experience but have worked in an election before, can be disorienting. P14 said, ``\textit{every year the screen} [to close the polls] \textit{is a little different, the lack of software consistency really throws people off}.'' Timing these updates, especially if a vulnerability is found close to election day, becomes a challenge for election officials. States and jurisdictions are sometimes forced to choose between theoretical election security and running a smooth election---a precursor to public trust that could keep election officials safe. Either deploy a system with a known vulnerability but consistent performance that everyone has trained on, both election officials and poll workers, or depend on every election official pushing a complicated update on an entire suite of election technology in a short period of time.

\subsubsection{\textbf{Election officials spend time catching mistakes and troubleshooting user error and malfunctions}}
\label{results:sub:burden} 

Even when some regions have the budget for extra technical support on election days, election officials reported that poll workers prefer to reach out to them first. ``\textit{If someone calls me, they'll feel more comfortable telling me or my clerk that they're having an issue, and then I'll reach out to the} [IT] \textit{technician},'' shared P10. However, when there is a lack of technical experience among poll workers, this creates a burden for election officials. ``\textit{Two years ago, the }[electronic pollbooks] \textit{ functionality, I felt like we kept having to explain it over and over again},'' shared P02. And sometimes the lack of technical experience means the problem a poll worker is seeking help for is not always a real problem: ``\textit{poll workers want something to work immediately... it's like, just give it a minute to register things are plugged in},'' shared P05. Sometimes, the problems are not technical, but psychological, and election officials find themselves reassuring poll workers that they are correctly using their technology. The technical support offered by election officials now becomes emotional support as they serve poll workers worried about making mistakes in a high-stakes environment. S37 shared that ``\textit{the poll workers, they get so stressed} [with electronic poll books]... \textit{take your time, slow down}.'' Despite the extra responsibility, many election officials sympathized with the poll workers who need extra help: ``\textit{when you first get something new, you're scared to use it, you know?}'' explained P05.  

However, managing these mistakes and errors is labor-intensive for election officials. First, election officials have to spend extra time being available as a resource and keeping track of poll worker activity to catch mistakes. Some election officials described poll workers having ``\textit{apprehension}'' towards reporting mistakes. ``\textit{I get that} [you're scared of making a mistake], \textit{just tell me what you're doing... I think sometimes people are so afraid to make a mistake or let somebody know they've made a mistake},'' shared P06. When a mistake does happen, a majority of election officials who reported seeing preventable mistakes while using election technology agreed that these mistakes cost a lot of time to fix or remedy (Fig. \ref{fig:fixingmistakes}). In the mistake P04 described in (\secref{results:sub:mistakes}), the fix required several steps and focused attention from that election official: 

\begin{quote}
    ``So, we couldn't read the results cards. I had to get the tabulator back on election night to turn it back on, put the cards in, and properly close the polls in order for us to get the results. It can affect prompt and accurate reporting of election results due to small, thoughtless measures of how they design the interface.'' ---\textit{P04}
\end{quote}

\subsubsection{\textbf{Election officials deal with consequences from voter distrust towards election technology}}
\label{results:sub:managedistrust}

Trust in elections among voters in the U.S. remains low \cite{saad2024partisan} and exists across both political parties \cite{kousser2026trust}. Public reporting \cite{staff_voting_2023, oosting_jonathan_human_2024, niesse_fulton_2024, harrington_election_2022} of mistakes like the ones mentioned in our earlier analysis (\secref{results:sub:mistakes}) erode public trust, even when they do not change the outcome of an election. Election officials in our survey mostly disagreed that mistakes when using election technology can cause a change in the final count (Fig. \ref{fig:fixingmistakes}), but a majority agreed that these types of mistakes can cause confusion with the public (Fig. \ref{fig:fixingmistakes}). Even with debunking, conspiracies about election technology remain popular enough to influence vendor reputations \cite{riccardi2025dominion}. 

As a result of this distrust, election officials act as unofficial crisis managers in their own local communities. They handle communications, anticipate rumors, and deal with conspiracies with voters they interact with personally. One participant shared that part of their job duty is to listen to radio shows for misinformation. When a caller started spreading a lie about election technology switching votes in local polling locations, they worked with the radio talk show host to dispel the rumor. ``\textit{If we didn't have relationships with the people in the community, that would have gotten out of control.}'' P04 described working with other election officials after 2020 to have ``\textit{everybody on the same page about saying what we do to protect the vote, why we are so sure that their vote was counted accurately} [with our systems].'' Election officials now feel the need to be spokespeople for their election technology.

Election officials had a keen sense of what types of technical problems could cause more misinformation. P06 had to ``\textit{run} [votes] \textit{through a new memory card on a different scanner}'' when a malfunction happened. ``\textit{Even something like that just creates space for someone to say you replaced it with a bad or fudged memory scanner or something like that}.'' A participant in a state where local jurisdictions choose their own technology said that they thought their state could never adjust to a single system because ``\textit{if there was a vendor every single county used, there would be people who claim that the tabulator isn't secure or whatever}.'' P09, a state election official, described a program where their team went to different regions to have forums with voters who ``\textit{don't trust elections or have questions about elections}.'' They described having an opening presentation with ``\textit{an explainer about voter list maintenance, mail ballots, and risk-limiting audits... those are the things that usually people are concerned most about}.''

When technical problems become public or when conspiracies target local areas, election officials personally deal with the fallout. P01 shared how when ``\textit{display errors were given out as actual data}'' in their voter registration database, ``\textit{that caused so many conspiracies}.'' P12's ``\textit{entire election staff}'' was ``\textit{wiped out... they were harassed by people accusing them of rigging the vote}.'' It was ``\textit{very nasty}.'' P14 said they have ``\textit{definitely seen the really nasty attitudes that get displayed towards election officials and poll workers}'' as a result of misinformation. In rural or suburban areas, navigating distrust towards election technology means potentially engaging with voters in person in their private lives:    

\begin{quote}
    ``Because it is a small area, people recognize you even if they don't necessarily know you. So, you'd have people coming up to you that you've never seen before in your life. `Hey, why are you doing blah, blah, blah. Why did you let your machines connect to the internet?' We don't let our machines connect to the internet. There's not a modem in our machines. Like they'll yell at you at Walmart and things like that.''---\textit{P06}
\end{quote}

The election officials we spoke to work hard to restore voter trust about election technology and often get marginal returns on those efforts. In smaller jurisdictions, a staff of one or two election officials might find themselves fighting against a nationwide conspiracy theory. Months of work can be dismantled by one mistake---``\textit{if you slip up, just one slip up, it's going to be this huge ordeal},'' according to P05. Burnout, fatigue, and a fear of seeking help due to the risk of causing more trouble were common among our participants. 

\begin{quote}
    ``I can't tell you how many times the night before an election I'm crying because I'm just so overwhelmed... or just the complete exhaustion... and then people just scrutinize. There's a certain group... they come to every board meeting and they scrutinize everything we do.''---\textit{P05}
\end{quote}

\subsubsection{\textbf{Election officials are not equipped to deal with personal digital safety risks}}
\label{results:sub:risk}
Election officials have become the targets of harassment and physical violence due to election misinformation and political polarization in the U.S. Sometimes, the tensions from voters explored in (\secref{results:sub:managedistrust}) as a result of misinformation and polarization extend to the digital realm. The election officials we spoke to had experienced doxxing, hacking, and misinformation about them online. However, despite these risks, election officials were generally less prepared to handle their \textit{personal digital} safety on their private devices and accounts. Instead, they were better equipped to deal with digital security threats at their offices and protect their physical safety. While these protections are still important, these personal digital threats can still lead to election official harm and security risks. 

Election officials generally received less advice in protecting their devices at home, especially when it comes to secure communication and account protection. ``\textit{Our work email... if I do have to send a secure email, I can... but personally, I don't have anything},'' one participant shared. Another participant had only adopted a password manager after something ``\textit{scary}'' happened to them in 2020. Lack of digital safety advice was often directly related to availability of IT support (\secref{results:sub:ITsupport}). ``\textit{When you have very little cybersecurity training, no program in place to test for phishing or anything like that, and you're bringing up a hotmail account, that's not good},'' P01 said. P03 shared a similar experience. ``\textit{Without having an IT department, it's real hard to do anything other than don't open attachments}.''

Many participants expressed wanting more protection for their personal digital safety, but several factors complicate this endeavor. FOIA requests allow contact information to be revealed about election officials, and that meant ``\textit{even if my personal life is completely secure, that doesn't matter because the public can come in, do a public records request and get any information they want about me}'' (P13).  
 
Election officials can be limited by the law when it comes to addressing a digital threat they have experienced in their personal capacity. P06 said their local law enforcement told them ``\textit{I can't arrest somebody based on a Facebook post}'' when poll workers were falsely accused and faced hate and harassment online. Having a contact to report digital threats to is itself a privilege---``\textit{you're going to find a lot of the counties that have a hard time getting the deputies to pick up the phone} [over digital threats] \textit{are also the ones who have poor IT departments},'' P02 shared.

\subsection{How U.S. Election Officials Overcome Challenges Related to Election Technology}
\label{results:overcome}

\begin{figure}
    \includegraphics[width=1\linewidth]{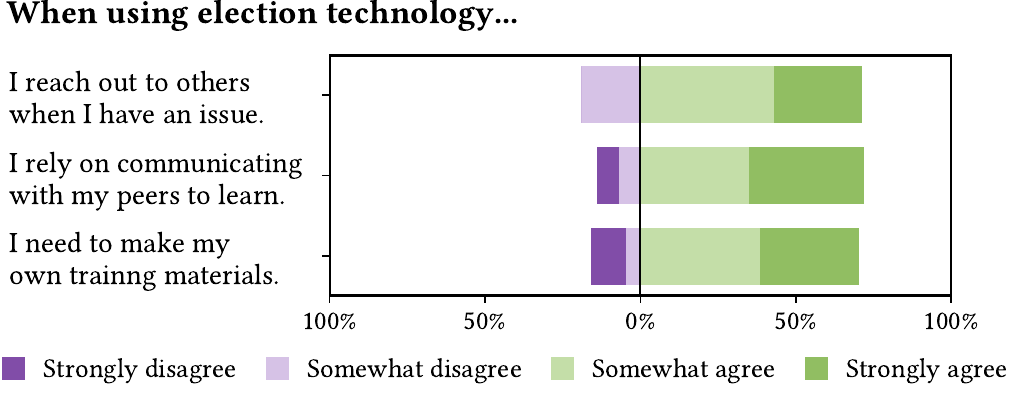}
    \caption{Likert-scale results of the 45 election officials who participated in the online survey. The left side of the chart shows disagreement and the right side shows agreement.}

    \label{fig:communitytech}
\end{figure}

\subsubsection{\textbf{Depending on institutional knowledge and self-made documentation}} 
\label{results:sub:documents}
In our survey, 80\% of election officials agreed that there is someone on their team that, if they left before the next election, would significantly impact their ability to run elections. For many of our election officials, it took years to develop an innate understanding of their operations. Two-thirds of our survey participants agreed that the things they need to know to do their job are often not on paper. P05 explained, ``\textit{there are just so many rules you have to follow to just do an election, and you can't learn it all in one election or two elections... you have to get thrown to the wolves a bit, and experience it, to learn it}.'' Combined with persistently low budgets and limited staff (\secref{results:sub:ITsupport}, \secref{results:sub:funding}), and the consequence is that election officials are often highly dependent on the institutional knowledge of one or two experienced colleagues, often themselves. 

For the election officials we spoke to, that dependence on institutional knowledge extends to using and understanding election technology. Documentation provided by the state and vendors is often not sufficient to overcome a lack of experience. P01 described their own journey sifting through the documents provided to them to understand their systems---``\textit{before I started working in elections, I thought the government had this plan, like a central database...that doesn't exist}.'' And with vendor documentation, P01 said that ``\textit{it tells you how to turn it on, but not when to turn it on or what security checks to run as you turn it on}.''

The solution for many election officials is to make their own documentation to record this institutional knowledge about election technology. In our survey, 68\% of election officials agreed they feel the need to make their own training materials for election technology (Fig. \ref{fig:communitytech}). Roughly 77\% of election officials said they had made or used documentation made by other election officials in their training. Of those election officials, 53\% agreed they could not complete their job without those materials (for P13, ``\textit{all the training materials are made in-house}''). Without this homemade documentation, the burden is placed on election officials to clear up discrepancies on election days. P14 shared, ``\textit{anytime there's paper documentation that doesn't match with reality...} [the poll workers] \textit{freak out}.'' However, creating documentation is a labor-intensive process. For some of our interview participants, the creation of these materials is a long-term goal or work-in-progress:

\begin{quote}
    ``Our voter registration system is probably the least usable thing we have but that's gotten better. [Another election official in the office] spends a lot of [their] time writing step-by-step instructions for how to do certain things or how to get around certain problems. There's a lot of workarounds for that system that aren't necessarily intuitive that add to the time and the amount of man hours it takes to do anything in that system.''---\textit{P01}
\end{quote}

\subsubsection{\textbf{Calling on strong communities and close relationships for technical help and support}} 
\label{results:sub:community}
The election officials who spoke with us revealed close-knit networks with other election officials in neighboring jurisdictions and across the U.S. They develop strong working relationships that can turn into lasting, personal friendships. Those connections allow them to seek advice about their election technology when they do not have access to their states or vendors, or when election workers want low-stakes interactions where they feel comfortable asking for help. In our survey, 68\% of election officials agreed that they rely on communicating with their peers to learn about election technology, and 66\% agreed they often reach out to others when encountering an issue with their election technology (Fig. \ref{fig:communitytech}),  like this participant: 
\begin{quote}
    ``The [election official] in the neighboring county to us, they actually worked here and just went in for a new job over there. So like, they and I text every day about, I mean, our kids, let alone just like, hey, what are you doing with this? Or, what are you doing with this piece of equipment?''---\textit{P05}
\end{quote}

Because of the desire to seek help directly from colleagues, election officials with more technical experience become pillars of the community. If an election official we spoke to was not that person in their community, they knew who was. ``\textit{There are three or four election directors across the state that are pretty tech savvy, and so I usually call them}'' for help, explained P06. For election officials we spoke to who do provide technology help and advice to others, they often take a passionate, proactive approach to providing guidance:
\begin{quote}
    ``So just teaching better document management, better password management... a lot of election offices are cut off from their main county. I always say it's like the little Harry Potter room under the stairs---they don't think about you unless it's election time. So helping them communicate with their county administration... for example, in one county, the internet was being beamed through a satellite dish through a thicket of pine... I was like, you need to talk to [the county] about getting a hardwire.''---\textit{P02}
\end{quote}

However, beneath these local relationships is a broader fear that technical problems to leak outside of regional communities. As discussed in (\secref{results:sub:risk}), election officials are acutely aware of public records requests, on top of feeling helpless when dealing with online narratives about their work. P02 noted that ``\textit{so many people are trying to have phone conversations}'' instead of email, which they said is ``\textit{unfortunate because there are certain things you're gonna need a paper trail for}.'' There is a trade-off between hyper-local, casual communications to deal with technology problems and broader transparency. P02 continued, ``\textit{there are a lot of things that people have gotten so scared about being open record requested for that I feel like should be in writing}.'' 

Another consequence of these communication practices is the ability to receive support from larger institutions like the state, who may have the ability to solve the technical problems they are facing. P11 explained how a state-level security response team they created had struggled to provide support because local counties were trouble-shooting issues before reaching out to them. 

\begin{quote}
    ``They are triaging internally and trying to figure this out before they ever call us. And [the security team] was like, why wouldn't they call us first? We're here to help them. We [the state] are like, that's just the culture, right? They will call us, but it's like, you know, shameful, like having to go to dad.'' ---\textit{P11}
\end{quote}

%% file: 6discussion.tex
\section{Discussion}
We have shown how challenges election workers have when using election technology cause real, measurable harms to elections. In our findings, we have reported examples of election technology causing longer wait times, preventing ADA machines from being deployed, allowing incomplete or erroneous records, preventing eligible voters from checking in, and causing mistakes that are hard for the public to understand. None of these examples changed the outcome of an election. However, these findings indicate a need to move beyond binary definitions of success when evaluating election technology. It is not enough to measure election technology as secure or not secure, accurate or not accurate. There is a spectrum of technology-related challenges to be addressed, each contributing to a general effect of harm for election workers and democracy.

In reducing election technology problems to whether or not they cause a final change in the election outcome, we ignore the consequences incidents have on election work through other negative effects. The election officials we spoke with were wary of problems with election technology not because they were worried about changes to the outcome, but because they understood how those problems escalate into losing public trust, inviting scrutiny, and fueling misinformation. That misinformation turns into direct threats to the security and safety of election workers, leading election workers to avoid help, burn out, and quit---creating an environment conducive to more mistakes. We argue that this negative feedback loop can be partially mitigated by improving election technology.

Our discussion covers reflections and recommendations to address the problems we found in common across various election technologies. Humans will make mistakes, and technology will fail. However, we explain why many of the challenges with election technology shared with us were caused by basic design flaws usability researchers have studied how to address. We argue that security research must consider human factors when creating protocols in order to design verifiable systems. In suggesting future work, we outline how the HCI community is uniquely positioned to address challenges with technology in U.S. elections, and provide potential solutions for improving collaboration and research.

\subsection{Improving Election Technology}
\label{discussion:improving}

\subsubsection{\textbf{Introduce defensive design standards}} This work extends Halderman's report of ``insufficiently defensive software design'' in Antrim County, Michigan \cite{haldermanAntrimCounty20202022} to nationally used election technology covered in our sample.  Future election technology should adopt frameworks used to situate human error as a side effect of system design or explicitly acknowledges humans-in-the-loop \cite{reasonHumanErrorModels2000, normanDesignRulesBased1983, cranorFrameworkReasoningHuman2008}. For example, we can show how election technology discussed in this work violated several conditions of Norman's lessons to design systems robust to human error \cite{normanDesignRulesBased1983}. First, that systems should provide \textit{feedback}. When an experienced poll worker selected cancel instead of proceed on an unfamiliar prompt, the interface did not provide feedback that the current button selection would cancel a voter's check-in record, not simply return the electronic pollbook back to the original state. Second, that actions should not have \textit{similarity} of response sequences---for instance, buttons that do different tasks should be clearly separated, like when a poll worker thought they closed the polls but had accidentally pressed the button that turned off the device. Third, that systems should be \textit{consistent} to reduce memory burden, although, with how decentralized election technology is, that may be impossible. Only \textit{reversibility} seems routinely adopted, given that the election official or poll worker catches the issue. In some cases, like the voter check-in example, reversibility has a time limit to act on. 

\subsubsection{\textbf{Find better solutions for accountability requirements}}

This work showed that election technology can be designed in a way that can overwhelm even experienced election workers, which also led to mistakes. Our participants shared how complicated accountability and auditing requirements determined by state governments (e.g., RLAs) can be reflected in the design of the system. In discussing the harms of making invisible work visible, Star and Strauss claim ``in the name of legitimacy and achieving public openness, an increased burden of accounting and tracking may be incurred'' \cite{starLayersSilenceArenas1999a}. This has happened to election workers, who face more pressure to be transparent and accountable in the presence of increasing public scrutiny. There is tension between needing information on election processes for integrity (e.g., audits \cite{lindemanGentleIntroductionRiskLimiting2012, PostElectionAudits}) and the extra data entry and documentation causing inconsistencies that undermine integrity.

\subsubsection{\textbf{Test for reliability}} The election officials we spoke to experience technology failures consistently. Some of these failures were a constraint on the technology itself, like rural precincts struggling to connect when there is low reception. Many were instances where election technology randomly stopped working, for example, when a memory card broke during election day. In addition to our findings, there is external validation that election technology can be unreliable. In 2022, Maricopa County, Arizona made national headlines when printer malfunctions occurred \cite{czopekWereMaricopaCounty2022}.  In 2023, as part of a legal settlement, the Commonwealth of Pennsylvania now requires election officials to fill out an incident report whenever election technology malfunctions on election day \cite{derosechristopherLawsuitSettlementLeads2023} (another ``increased burden of accounting'' \cite{starLayersSilenceArenas1999a}). Failures were documented in several reports from 2024, including corrupted metadata that had ``never been seen before'' and that forced a manual re-scanning of ballots when election officials could not figure out the cause of the error.  

During an election cycle, electronic pollbooks, tabulators, and other election equipment will be used thousands of times over the span of several weeks. Even a low incidence rate of election technology failure can be consequential. These failures have downstream effects like creating longer lines at polling centers and delaying voter check-ins, threatening enfranchisement. As a result, some election officials exercise caution when trusting election technology to work reliably and advised trainees to do the same. While exercising more caution is never a bad idea, remaining vigilant to catch failures does not prevent the entirety of their negative effect. Election technology should be stress-tested more rigorously and testing results shared with election officials and state governments before making purchases.  

\subsubsection{\textbf{Incorporate usability into evidence-based protocols}}
Our findings suggest that when election systems are  error-prone, critical assumptions about election verification methods are easily violated. This is because proposed solutions for verification rely on manual configurations provided by election officials (e.g., \cite{benalohElectionGuardCryptographicToolkit2024}), and currently enacted auditing methods like risk-limiting audits require complete and accurate ballot records \cite{lindemanGentleIntroductionRiskLimiting2012}. Evidence-based election methods cannot provide their original guarantee using inaccurate data. Our participants  and public examples (e.g., \cite{haldermanAntrimCounty20202022, starkWhenAuditsRecounts2024}) have shown how human error  caused by interface design or usability problems lead to election workers submitting the wrong information. Usability and verification have a stronger dependency than is reflected in the implementation of these methods. Until solutions are found to improve the design of election technology, security research should take user requirements into more consideration.

\subsection{Future Directions to Support Election Work}
\label{discussion:future}

\subsubsection{\textbf{Prioritize usability research for election workers}} Future research should consider the usability of election systems as a necessary requirement for election integrity and election worker safety. Local, in-depth case studies of specific jurisdictions and how they use their election technology would provide a foundation to address usability failures for specific vendor technologies (e.g., tabulators or electronic pollbooks). Participatory design sessions, for example, could allow researchers to identify where issues are most common across different jurisdictions. To the best of our knowledge, a publicly available, peer-reviewed or independent cognitive walkthrough or usability study of an actively deployed worker-facing election technology does not exist. Alternatively, usable security researchers can focus on a single integrity protocol (e.g., RLAs) and see where usability flaws most frequently threaten the protocol's security assumptions. Or, HCI researchers could design their own election technology back-ends, much like they designed ballot casting front-ends in the past. These studies could apply further pressure on vendors to support defensive and simplified election technology design. 

\subsubsection{\textbf{Consider the digital safety of election workers}} The growing field of usable privacy and security is aptly suited to address the unique security needs of election workers. Future work should expand upon our introductory findings about the digital safety practices and needs of election workers. As critical stakeholders in a democratic system, election workers' safety is integral to free and fair elections. Computer scientists and digital safety researchers should partner with local election officials, voting non-profits, and other stakeholders to consider organizing support systems for election workers, like digital safety clinics and trainings.

\subsubsection{\textbf{Build trust with the elections community}} To meaningfully engage with election officials, computer scientists must build trust. As an at-risk group subject to harassment, election officials might be hesitant to engage with academics. In addition to that, there are preexisting tensions between election officials and computer scientists. In our survey, a participant responded saying they were ``somewhat comfortable'' working with computer scientists, and when elaborating why or why not, they responded, ``[A named security researcher]? \textit{Nope}.'' In the context of our survey, we asked this question in the same section we ask participants if they would be interested in conducting an interview---it is likely that this participant meant they were interested in engaging with this project as long as this security researcher was not involved. To build better relationships between election officials and computer scientists, researchers should signal from the beginning that they want to work with election officials. Work should come from a place of understanding---we show how election officials are running elections while dealing with technical constraints, resource limitations, and threats of harm. Future researchers should find trusted advocates to make introductions to local election communities, following best practices recommended in HCI and CSCW research with at-risk groups.

\subsubsection{\textbf{Encourage  protections for research and collaboration}} Research on U.S. election technology is limited by the lack of opportunities for election officials to engage in research, and states and election vendors have avoided interaction with researchers in the wake of election misinformation. There are several policy ideas that researchers can champion to improve the viability of research in U.S. elections: require the sharing of data about technology malfunctions during active elections, allow academics to observe election officials and poll workers in the field (with legal protections), and make election technology more accessible and, when applicable, open source. 

Some of these recommendations are already in process, though to a limited extent. As discussed, in 2023, the Commonwealth of Pennsylvania now requires election officials to fill out an incident report whenever election technology malfunctions on election day \cite{derosechristopherLawsuitSettlementLeads2023}. We presented this requirement as another ``burden of accounting'' to legitimize election integrity \cite{starLayersSilenceArenas1999a}. The data from 2024 reaffirms that conclusion---upon inspection, many forms are not entirely filled out or cover multiple incidents in a single report. Researchers need this information to identify problem areas in election technology, but this data must be complete, accurate, and trustworthy. An initiative like this should incorporate trusted, nonpartisan third-party observers to reduce the burden on election officials and poll workers. 

Researchers should encourage public officials to provide protections to study the entire election process with election officials and poll workers in experimental settings. This is a tough request when current computer scientists scarcely have access to isolated machines, let alone access to an entire system or open-source election technology code. But, our findings prove the importance of studying an election system as a whole.

%% file: 7conclusion.tex
\section{Conclusion}
\label{conclusion}

This paper shows that election officials face consistent problems across vendors and jurisdictions when leveraging election technology in their work. While our recommendations focus on which challenges HCI research can address, our findings imply other necessary improvements are needed. Election officials must have adequate technical support and sufficient funding for technology upgrades and training. Technology problems and resources are not separable in election work---the jurisdictions with some of the toughest challenges related to technology are the ones that have the least support.

These technical challenges lead to preventable mistakes, delayed security updates, and additional burdens that election officials absorb outside of their normal job responsibilities. Election officials recognize a direct relationship between the challenges they have with election technology and the harms they face---mistakes and failures feed public distrust or misinformation, and misinformation turns into harassment and personal safety risk many election officials felt unequipped to handle. Their understanding of this connection impacts how election officials resolve and disclose future challenges with election technology. For example, Moran et al. found that FOIA requests led to paranoia and self-censorship \cite{moranPrivacyTransparencyNavigating2025}, a finding echoed by our participants with potential consequences for seeking help and resolving errors and security threats.   

To overcome these challenges, election officials rely on local communities where they feel safe and supported. These communities become resources as election officials make homemade documentation and guides to offset confusing election technology or user manuals. There is a strong dependency on institutional knowledge, especially when overcoming technical challenges. 

We argue that there are future directions for researchers to help assist election work and reduce the technical burden presented in our findings. That is because some of these problems---like insufficiently defensive design---can be measured, tested, and improved. While our proposals will never prevent election misinformation that is completely fabricated, or lighten the emotional weight of operating an election, supportive election technology can ease some of the symptoms described to us by election officials. Over time, the improvements mentioned in this work could increase trust in elections, bolster election security, and protect election workers.

%% file: 8appendix.tex
\appendix

\section{Survey Statistics}

\textbf{1. I feel confident in poll workers’ ability to troubleshoot our election technology during an election when I am not present} (\secref{results:sub:support}).

\begin{tabular}[t]{|l|c|}
\hline
Strongly Agree &  4 \\ \hline
Somewhat Agree & 15 \\ \hline
Neither Agree nor Disagree & 5 \\ \hline
Somewhat Disagree & 12 \\ \hline
Strongly Disagree & 8 \\ \hline
No Answer & 1 \\ \hline
\end{tabular}
\newline 
\newline

\textbf{2. When election officials receive new election technology or when election technology is updated, it takes a while to understand even with instructions.} (\secref{results:sub:support}).

\begin{tabular}[t]{|l|c|}
\hline
Strongly Agree &  7 \\ \hline
Somewhat Agree & 18 \\ \hline
Neither Agree nor Disagree & 7 \\ \hline
Somewhat Disagree & 9 \\ \hline
Strongly Disagree & 3 \\ \hline
No Answer & 1 \\ \hline
\end{tabular}
\newline 
\newline

\textbf{3. Others making mistakes while using election technology are likely enough to where I’m stressed or anxious about them during an election.} (\secref{results:sub:mistakes}, Fig. 1).

\begin{tabular}[t]{|l|c|}
\hline
Strongly Agree &  6 \\ \hline
Somewhat Agree & 13 \\ \hline
Neither Agree nor Disagree & 7 \\ \hline
Somewhat Disagree & 6 \\ \hline
Strongly Disagree & 12 \\ \hline
No Answer & 1 \\ \hline
\end{tabular}
\newline 
\newline

\textbf{4. Election technology requires oversight to make sure mistakes aren’t made.} (\secref{results:sub:mistakes}, Fig. 1).

\begin{tabular}[t]{|l|c|}
\hline
Strongly Agree &  21 \\ \hline
Somewhat Agree & 12 \\ \hline
Neither Agree nor Disagree & 5 \\ \hline
Somewhat Disagree & 5 \\ \hline
Strongly Disagree & 1 \\ \hline
No Answer & 1 \\ \hline
\end{tabular}
\newline 
\newline

\textbf{5. It’s common for new poll workers to make mistakes when using election technology.} (Fig. 1). %

\begin{tabular}[t]{|l|c|}
\hline
Strongly Agree &  11 \\ \hline
Somewhat Agree & 18 \\ \hline
Neither Agree nor Disagree & 7 \\ \hline
Somewhat Disagree & 3 \\ \hline
Strongly Disagree & 5 \\ \hline
No Answer & 1 \\ \hline
\end{tabular}
\newline 
\newline

\textbf{6. I have seen preventable mistakes made while using election technology during an election.} (\secref{results:sub:mistakes}). %

\begin{tabular}[t]{|l|c|}
\hline
Yes &  33 \\ \hline
No & 11 \\ \hline
No Answer & 1 \\ \hline
\end{tabular}
\newline 
\newline

\textbf{6A. These mistakes could cause confusion with the public.} (Of the 33 who answered ``yes'' to 6.) (\secref{results:sub:managedistrust}, Fig. 2).

\begin{tabular}[t]{|l|c|}
\hline
Strongly Agree &  15 \\ \hline
Somewhat Agree & 12 \\ \hline
Neither Agree nor Disagree & 3 \\ \hline
Somewhat Disagree & 2 \\ \hline
Strongly Disagree & 1 \\ \hline
\end{tabular}
\newline 
\newline

\textbf{6B. These mistakes cost a lot of time to fix or remedy.} (Of the 33 who answered ``yes'' to 6.) (\secref{results:sub:managedistrust}, Fig. 2).

\begin{tabular}[t]{|l|c|}
\hline
Strongly Agree & 4 \\ \hline
Somewhat Agree & 16 \\ \hline
Neither Agree nor Disagree & 3 \\ \hline
Somewhat Disagree & 5 \\ \hline
Strongly Disagree & 5 \\ \hline
\end{tabular}
\newline 
\newline

\textbf{6C. These mistakes could cause a change or error in final counts.} (Of the 33 who answered ``yes'' to 6.) (\secref{results:sub:managedistrust}, Fig. 2).

\begin{tabular}[t]{|l|c|}
\hline
Strongly Agree &  3 \\ \hline
Somewhat Agree & 4 \\ \hline
Neither Agree nor Disagree & 2 \\ \hline
Somewhat Disagree & 11 \\ \hline
Strongly Disagree & 13 \\ \hline
\end{tabular}
\newline 
\newline

\textbf{7. There is someone on our team that, if they were to leave before the next election, would significantly impact our ability to run elections.} (\secref{results:sub:documents}). %

\begin{tabular}[t]{|l|c|}
\hline
Strongly Agree &  19 \\ \hline
Somewhat Agree & 17 \\ \hline
Neither Agree nor Disagree & 1 \\ \hline
Somewhat Disagree & 5 \\ \hline
Strongly Disagree & 1 \\ \hline
No Answer & 2 \\ \hline
\end{tabular}
\newline 
\newline

\textbf{8. Things I need to know to do my job are often not documented or on paper.} (\secref{results:sub:documents}). %

\begin{tabular}[t]{|l|c|}
\hline
Strongly Agree &  17 \\ \hline
Somewhat Agree & 13 \\ \hline
Neither Agree nor Disagree & 5 \\ \hline
Somewhat Disagree & 4 \\ \hline
Strongly Disagree & 4 \\ \hline
No Answer & 2 \\ \hline
\end{tabular}
\newline 
\newline

\textbf{9. I feel the need to make my own training materials that explain how to use our election technology.} (\secref{results:sub:documents}). %

\begin{tabular}[t]{|l|c|}
\hline
Strongly Agree &  16 \\ \hline
Somewhat Agree & 13 \\ \hline
Neither Agree nor Disagree & 7 \\ \hline
Somewhat Disagree & 6 \\ \hline
Strongly Disagree & 2 \\ \hline
No Answer & 1 \\ \hline
\end{tabular}
\newline 
\newline

\textbf{10. I have made instructional media or used instructional media made by other election administrators to help with operations (e.g., a Canva poster).} (\secref{results:sub:mistakes}). %

\begin{tabular}[t]{|l|c|}
\hline
Yes &  35 \\ \hline
No & 9 \\ \hline
No Answer & 1 \\ \hline
\end{tabular}
\newline 
\newline

\textbf{10A. I could not complete my job without these instructional media to help with operations (e.g., a Canva poster).} (Of the 35 who answered ``yes'' to 10.) (\secref{results:sub:documents}). %

\begin{tabular}[t]{|l|c|}
\hline
Strongly Agree &  17 \\ \hline
Somewhat Agree & 7 \\ \hline
Neither Agree nor Disagree & 6 \\ \hline
Somewhat Disagree & 4 \\ \hline
Strongly Disagree & 1 \\ \hline
\end{tabular}
\newline 
\newline

\textbf{11. I rely on communicating with my peers (other election administrators, even in other counties) to learn how to use election technology.} (\secref{results:sub:community}, Fig. 3). %

\begin{tabular}[t]{|l|c|}
\hline
Strongly Agree &  16 \\ \hline
Somewhat Agree & 15 \\ \hline
Neither Agree nor Disagree & 6 \\ \hline
Somewhat Disagree & 3 \\ \hline
Strongly Disagree & 3 \\ \hline
No Answer & 2 \\ \hline
\end{tabular}
\newline 
\newline

\textbf{12. I often have to reach out to others when encountering an issue with election technology.} (\secref{results:sub:community}, Fig. 3). %

\begin{tabular}[t]{|l|c|}
\hline
Strongly Agree &  12 \\ \hline
Somewhat Agree & 18 \\ \hline
Neither Agree nor Disagree & 4 \\ \hline
Somewhat Disagree & 8 \\ \hline
Strongly Disagree & 0 \\ \hline
No Answer & 3 \\ \hline
\end{tabular}
\newline 
\newline

\textbf{13. I feel the need to make my own training materials to explain how to use our election technology.} (Fig. 3). %

\begin{tabular}[t]{|l|c|}
\hline
Strongly Agree &  14 \\ \hline
Somewhat Agree & 17 \\ \hline
Neither Agree nor Disagree & 6 \\ \hline
Somewhat Disagree & 2 \\ \hline
Strongly Disagree & 5 \\ \hline
No Answer & 1 \\ \hline
\end{tabular}
\newline 
\newline

\section{Interview Outline}

\subsection{Role and Technology Use}

\begin{itemize}
    \item What is the name of your title? What are your main responsibilities?
    \item Who are some people that you work with? How are responsibilities distributed?
    \item How often do you interact with elections-specific equipment (registration systems, scanners, counters, election night reporting systems)?
    \item What other technologies do you use in your work? What for? 
\end{itemize}

\subsection{Training Election Workers}

\begin{itemize}
    \item Who is responsible for training on various tasks?
    \item When you are training new election officials or poll workers, how do you walk them through technology concepts? How do you walk them through the laws related to election administration? How do they conceptualize the laws / how well do they understand them?
    \item Are there any new materials you’ve developed for training purposes? Have you shared those materials? 
    \item When training (other officials/poll workers), what struggle often comes up? What takes them the longest to learn? Can you provide an example?
\end{itemize}

\subsection{Failures with Technology Use}
\begin{itemize}
    \item Have you ever had an election official (including yourself) or poll worker fail to interact with election technology? (E.g., scanners, poll-pads, RLAs.) What was the mistake, and how did it happen? 
    \item What is one piece of election technology you would change, and how? 
    \item When you get stuck with a technical problem, who do you go to? Do you feel confident troubleshooting on your own? 
    \item Have you been wrongly accused of making mistakes? What was the accusation?
\end{itemize}

\subsection{Community and Practice}
\begin{itemize}
    \item How often do you engage with state employees? What is communication like? What do you communicate about? How do you think that affects your practices?
    \item What resources (e.g. training, learning) does the state have for counties? Anything security or safety oriented?
    \item How often do you share information with other counties? What is communication like? What do you communicate about? How do other counties run their elections based on your knowledge?
    \item Have you had experience with issues around digital personal safety? What are they? What happened? What protections or resources are given to you from outside your county? 
    \item Do you ever talk about security in the workplace? How often? What starts these conversations? Do you find them helpful? 
    \item Any other comments / stories you’d like to share today?
\end{itemize}